\documentclass[prd,twocolumn,amsmath,amssymb,floatfix,superscriptaddress]{revtex4-1}
\usepackage{graphicx}
\usepackage{amssymb}
\usepackage{amsmath}
\usepackage{color}
\usepackage{ wasysym }
\usepackage{hyperref}
\usepackage{tabu}
\usepackage{bm}
\usepackage{float}
\def\barray{\begin{array}}
\def\earray{\end{array}}
\def\be{\begin{equation}}
\def\ee{\end{equation}}
\def\ben{\begin{equation} \nonumber}
\def\een{\end{equation}}
\def\ban{\begin{eqnarray*}}
\def\ean{\end{eqnarray*}}
\def\ba{\begin{eqnarray}}
\def\ea{\end{eqnarray}}

\def\({\left(}
\def\){\right)}

\graphicspath{{./fig/}}

\begin{document}

\title{How Complex is Dark Energy? A Bayesian Analysis of CPL Extensions with Recent DESI BAO Measurements.}
\author{Mohammad Malekjani}
\email{malekjani@iasbs.ac.ir}
\affiliation{Department of Physics, Bu-Ali Sina University, Hamedan
	65178, Iran}
    \affiliation{Department of Physics, Institute for Advanced Studies in Basic Sciences (IASBS), P.O. Box 45137-66731, Zanjan, Iran}
\author{Saeed Pourojaghi}
\affiliation{School of Physics, Institute for Research in Fundamental Sciences (IPM), P.O.Box 19395-5531, Tehran, Iran}
\author{Zahra Davari}
\affiliation{School of Physics, Korea Institute for Advanced Study (KIAS), 85 Hoegiro, Dongdaemun-gu, Seoul, 02455, Korea}
\begin{abstract}
The nature of dark energy is one of the big puzzling issues in cosmology. While $\Lambda$CDM provides a good fit to the observational data, evolving dark energy scenarios, such as the CPL parametrization, offer a compelling alternative. In this paper, we present a Bayesian model comparison of various dark energy parametrizations using a joint analysis of Cosmic Microwave Background data, DESI Baryon Acoustic Oscillation measurements, and the PantheonPlus (or Union3) Supernovae type Ia sample. We find that while the $\Lambda$CDM model is initially favored over a constant $w$CDM model, the CPL parametrization is significantly preferred over $w$CDM, reinforcing recent evidence for an evolving dark energy component, consistent with DESI collaboration findings. Crucially, when testing higher-order CPL extensions, the so-called CPL$^+$ and CPL$^{++}$, our Bayesian analysis shows that the observational data do not favor these more complex scenarios compared to the standard CPL. This result indicates that adding excessive complexity to the CPL form is unwarranted by current observations. Interestingly, similar to the CPL parametrization, alternative two-parameter forms, specifically $w_{de}(a) = w_0 + w_b(1-a)^2$ and $w_{de}(a) = w_0 + w_c(1-a)^3$, yield a better fit to observational data than the standard $\Lambda$CDM cosmology. Our results challenge the necessity for overly complex CPL extensions and confirm that well-chosen two-parameter $w_0w_a$ parametrizations effectively capture DE evolution with current cosmological data, supporting the recent signals for dynamical dark energy by DESI collaboration.
\end{abstract}
\maketitle

\section{Introduction} \label{sec:intro}
The discovery of the accelerated expansion of the Universe, primarily inferred from Supernovae type Ia (SN Ia) observations nearly a quarter-century ago \cite{Riess:1998cb,Perlmutter:1998np}, revolutionized our understanding of the dynamics of the Universe. This paradigm-shifting finding required the introduction of a dominant, mysterious component, dubbed dark energy (DE), which constitutes approximately $68\%$ of the total energy density of the Universe \cite{Planck:2019nip}. Subsequent independent cosmological probes have robustly confirmed this acceleration and provided increasingly tight constraints on DE properties. Measurements of the Cosmic Microwave Background (CMB) anisotropies, particularly from missions like WMAP \cite{WMAP2013} and Planck \cite{Planck:2013pxb,Planck:2018vyg,Planck:2019nip}, give us a detailed picture of the early Universe and its composition, and also indirectly support the need for a DE component to reconcile geometry with observed matter densities. Furthermore, the large-scale distribution of galaxies, traced through the Baryon Acoustic Oscillations (BAO) in cosmological surveys like SDSS, BOSS, eBOSS, and most recently the DE Spectroscopic Instrument (DESI), provides a cosmic ruler that independently measures the expansion history, consistently referring to the cosmic acceleration \cite{Eisenstein:2005su,Alam2017, DES:2021wwk,DESI:2024mwx,DESI:2025zgx}. Cosmological data from galaxy clusters and weak lensing surveys support this picture, affirming that the accelerated expansion scenario is a cornerstone of our current cosmological frameworks \cite{Vikhlinin2009, Heymans2013}.\\
The simplest and most widely accepted explanation of the dynamics of the Universe on large scales is the $\Lambda$CDM model, where DE is identified with a cosmological constant ($\Lambda$). This model provides a perfect fit to the vast array of cosmological observations, such as high-resolution CMB anisotropies \cite{WMAP2013,Planck:2013pxb,Planck:2018vyg,Planck:2019nip}, the previous robust BAO measurements from large-scale galaxy surveys \cite{BOSS2017BAO}, and the latest extensive SN Ia data, such as the PantheonPlus, Union3 and the Dark Energy Survey five-year (DES-5YR) samples \cite{Scolnic:2021amr, Rubin:2023ovl,DES:2024tys}.\\
All these successes in describing the evolution of the early Universe and the present-day Universe have made the $\Lambda$CDM model the standard paradigm in cosmology. However, despite its observational successes, the $\Lambda$CDM model faces significant theoretical and some burgeoning observational challenges. Theoretically, the cosmological constant suffers from profound issues such as the "fine-tuning problem" and the "cosmic coincidence problem" \cite{Weinberg:1988cp,Padmanabhan:2002ji}. The former indicates the enormous discrepancy (of $\sim 120$ orders of magnitude) between the vacuum energy density predicted by quantum field theory and the amount we need to explain the observations. The latter problem is why DE has only become dominant, precisely when structure formation is occurring, instead of much earlier or later in the history of the Universe. From the observational point of view, while the $\Lambda$CDM model remains highly successful, persistent tensions have emerged between parameter values derived from different datasets. Most notably, a significant discrepancy exists in the inferred value of the Hubble constant ($H_0$), so that the early-Universe probes like the Planck experiment yield lower values than local measurements from the cosmic distance ladder \cite{Riess2019,Riess:2021jrx, Freedman2021}. Furthermore, some analyses show a $\sigma_8$ tension, where the amplitude of matter fluctuations at present is lower than the predicted value by CMB observations under the standard $\Lambda$CDM framework \cite{Pergola2023Tensions}.\\
We refer to a recent review \cite{Perivolaropoulos:2021jda} to address the observational tensions within the standard $\Lambda$CDM cosmology. Persistent challenges within the standard model have prompted cosmologists to consider frameworks that go beyond the conventional $\Lambda$CDM cosmology. Two main directions have emerged in this pursuit. One explores the possibility that DE evolves, exhibiting enough negative pressure while remaining consistent with general relativity. The other takes a more radical approach by suggesting modifications to general relativity itself, as encapsulated in various modified gravity theories (for a comprehensive review, see \cite{Tsujikawa:2010zza}). Early investigations into time-varying energy densities laid important groundwork for the idea of dynamical DE can be found in the seminal contributions by \cite{Copeland:2006wr,Caldwell:1997ii,Armendariz-Picon:2000ulo,Caldwell:1999ew,Padmanabhan:2002cp}. The time-dependent behavior of DE suggests the potential to alleviate cosmological tensions by allowing DE to become dominant at a specific epoch. Crucially, in the absence of a unique fundamental theory beyond the cosmological constant to describe the precise nature and evolution of DE, cosmologists typically resort to phenomenological parameterizations for the equation of state of DE, $w_{de}(z)$.\\
These phenomenological forms aim to capture the possibility for potential deviations from $w_{\Lambda}=-1$ and its time variability, without assuming a specific underlying cosmological model. Among these, the CPL (Chevallier-Polarski-Linder) parametrization, $w_{de}(a) = w_0 + w_a(1-a)$, stands out as a widely adopted and relatively simple parametrization, first introduced by \cite{Chevallier:2000qy,Linder:2002et}. It presents two free parameters, $w_0$ and $w_a$, allowing for linear evolution of the equation of state parameter of DE with the scale factor $a$. The CPL parametrization can capture a broad range of DE behaviors (including crossing the phantom divide $w=-1$), and it is usually the first choice for cosmologists aiming to measure the possible deviations from the cosmological constant. In addition to the CPL parametrization, various phenomenological parametrizations have been proposed. These include power-law forms such as $w_{de}(z) = w_0 + w_a \frac{z}{(1+z)^2}$ \cite{Jassal:2004ej} and $w_{de}(a) = w_0 + \frac{w_a (1 - a^\beta)}{\beta}$ \cite{Barboza:2009ks}, as well as logarithmic models like $w_{de}(a) = w_0 + w_a \ln{a}$ \cite{Efstathiou:1999tm} and $w_{de}(z) = w_0 [1 + b \ln{(1+z)}]^\alpha$ \cite{Wetterich:2004pv}. Each of these parametrizations offers distinct advantages and limitations. Moreover, the Padé parametrizations have been introduced \cite{Rezaei:2017yyj}, as a rational generalization of CPL. To explore further deviations and test the sufficiency of the CPL parametrization, one can naturally extend the linear CPL parametrization by including the higher-order terms in the Taylor series. In this regard, one can define two extensions as \textit{CPL$^+$}, which adds a quadratic term in $(1-a)$ to CPL as $w_{de}(a) = w_0 + w_a(1-a) + w_b(1-a)^2$ \cite{Linder:2005ne,SDSS:2004kqt} and \textit{CPL$^{++}$} which includes further extension by considering a cubic term ($w_{de}(a) = w_0 + w_a(1-a) + w_b(1-a)^2 + w_c(1-a)^3$) to describe the more complex evolution of the equation of state of DE. These parametrizations introduce more flexibility to the dynamics of DE  \cite{Nesseris:2025lke}.
These extended forms enable a more detailed examination of the time evolution of DE, potentially revealing features that simpler parametrizations may overlook. However, they introduce additional parameters that observational data must constrain. While the CPL parametrization and its higher-order Taylor extensions provide a systematic way to probe possible deviations from a constant equation of state $w_{\Lambda}=-1.0$, it is also useful to assume alternative two-parameter forms of equation of state that explore different shapes of evolution while maintaining a similar degree of flexibility to CPL. Specifically, we investigate parametrizations where the leading-order Taylor term beyond $w_0$ is not linear in $(1-a)$, but quadratic or cubic. This allows us to examine the impact of different forms of deviation from a constant equation of state, focusing on whether a specific non-linear shape of equation of state can be favored by the data, even if it has the same number of free parameters as CPL. We introduce two such parametrizations as: \textit{Quadratic $w_0w_b$ parametrization}, where we replace the linear $(1-a)$ term of CPL with a quadratic term as $w_{de}(a) = w_0 + w_b(1-a)^2$, maintaining two free parameters, and \textit{Cubic $w_0w_c$ parametrization}, where the linear term is replaced with a cubic term as $w_{de}(a) = w_0 + w_c(1-a)^3$, again maintaining two free parameters. These parametrizations offer an alternative perspective on the evolution of DE by eliminating the lower-order terms in the Taylor series and directly exploring nonlinear dependencies on $(1-a)$. By comparing these two-parameter shapes of the equation of state with CPL and its extensions (e.g., CPL$^+$ and CPL$^{++}$), our analysis aims to determine whether the cosmological data indicate a preference for a specific functional form of the equation of state of DE, by simply adding more terms to a linear expansion. We note that such an analysis could be valuable for determining the potential shape of $w_{de}(a)$ and for drawing the map for future theoretical investigations.\\
Recent studies in large-scale structure surveys have provided new insights into the dynamics of DE. DESI collaboration, through precise measurements of the clustering of galaxies, quasars, and the Lyman-$\alpha$ forest across a vast area of the sky and a wide range of redshift, has determined the transverse comoving distance and the Hubble rate in seven distinct redshift bins \cite{DESI:2024uvr,DESI:2025zgx}.
Furthermore, the full-shape analyses of the power spectrum of galaxies and quasars extended beyond the BAO scale in the DESI experiment, leveraging redshift-space distortions (RSD) to extract useful information on the growth rate of the large-scale cosmic structures, represented by $f\sigma_8$ \cite{DESI:2024jxi}. While the DESI BAO measurements, when analyzed alone, are generally consistent with a flat-$\Lambda$CDM cosmology, the combination of DESI BAO with CMB anisotropy measurements \cite{Planck:2018vyg} and different observations from SNIa samples (such as PantheonPlus \cite{Scolnic:2021amr}, Union3 \cite{Rubin:2023ovl}, and DES-SN5YR \cite{DES:2024tys}) reveals an interesting and groundbreaking feature of DE. In particular, recent statistical analyses by the DESI collaboration (e.g., \cite{DESI:2024mwx,DESI:2025zgx}), adopting the CPL parametrization for the equation of state of DE, indicate significant potential for the evolution of DE, showing considerable deviations from a constant $w_{\Lambda}=-1$. These compelling results, particularly the hints for evolving DE from DESI experiments, emphasize the essential need for a rigorous examination of the functional form of $w_{de}(a)$. In this regard, a recent Bayesian analysis \cite{Ong:2025utx} finds that, once the Bayesian evidence method is applied, the combination of DESI BAO Data Release 2 (DR2) and Planck CMB observations favors the $\Lambda$CDM cosmology over evolving DE models. However, when DES-5YR supernovae sample is included, the analysis indicates a preference for evolving DE.\\
In this paper, we conduct a detailed Bayesian statistical comparison of the standard CPL parametrization, its higher-order Taylor series extensions (CPL$^+$ and CPL$^{++}$), and two alternative two-parameter Quadratic and Cubic shapes. Our goal is to determine which of these specific parametrizations has the best support by the latest cosmological datasets, including DESI BAO (DR2), CMB anisotropies measurements, and SNe Ia observations, thereby providing crucial insights into the favored evolution of DE. The investigation into CPL extensions is particularly related to a recent analysis by \cite{Nesseris:2025lke}, where the authors studied higher-order CPL extensions (CPL$^+$ and CPL$^{++}$). Employing a frequentist statistical approach based on the standard Maximum Likelihood estimation (MLE)\footnote{For the data errors normally distributed and independent, the Maximum of Likelihood function is mathematically equivalent to the Minimum of $\chi^2$ function} and associated confidence regions, they utilized a combined dataset comprising CMB anisotropies, DESI BAO Data Release 1 (DR1) measurements, and the PantheonPlus SN Ia sample. Their findings indicate that, for both the CPL$^+$ and CPL$^{++}$ parametrizations, the posterior distributions in the $w_0$-$w_a$ parameter space are consistent with the $\Lambda$CDM point ($w_0=-1, w_a=0$). This result was interpreted as evidence supporting the $\Lambda$CDM model over these more complex CPL extensions, suggesting great care when using the CPL parametrization (e.g., \cite{DESI:2024mwx,DESI:2025zgx}).
Building on this context, we aim to perform a rigorous statistical test, utilizing a robust BE approach, to study the observational support for parametrizations extending beyond the standard CPL parametrization. In addition, we apply the statistical analysis based on the MLE and compare the two approaches. \\
The structure of the paper is organized as follows: In Section (\ref{sect:data}), we present the observational data used in this work. Section (\ref{sect:method}) introduces the statistical tools used in our analysis.  In Section (\ref{Sect:Numerical_results}), we present our numerical results for different DE parameterizations considered in our analysis. Finally, in Section (\ref{conlusion}), we summarize our findings and conclude the study.

\section{Observational data} \label{sect:data}
In this work, we utilize a combination of recent cosmological observations to constrain the model parameters. Specifically, we adopt the second-year BAO measurements from DESI \cite{DESI:2025zgx}, the Planck CMB data \cite{Planck:2018vyg} in the form of compressed distance measurements at recombination, also known as CMB shift parameters \cite{Zhai:2018vmm}, and SN Ia luminosity distance measurements from the Union3 \cite{Rubin:2023ovl} and PantheonPlus \cite{Scolnic:2021amr} compilations. The details of these datasets and their implementation in our analysis are discussed in the following subsections.

\subsection{DESI BAO}
The BAOs serve as a standard ruler to trace the cosmic expansion history through the characteristic comoving sound horizon at the drag epoch, $r_d$. The apparent BAO scale is measured in both the transverse and line-of-sight directions, providing the observables
\begin{equation}
    D_M(z)/r_d, \quad D_H(z)/r_d = \frac{c}{H(z)r_d},
\end{equation}
and their isotropic combination
\begin{equation}
    D_V(z)/r_d = \frac{1}{r_d}\left[z\, D_H(z)\, D_M^2(z)\right]^{1/3}.
\end{equation}
Here, $D_M(z) = (1+z)D_A(z)$ is the comoving angular diameter distance, and $D_H(z)$ is the Hubble distance.
In this work, we use the DESI BAO (DR2) dataset, based on the measurements summarized in Table~\ref{Tab:DR2_data}. The sample includes $D_V/r_d$ from the Bright Galaxy Sample (BGS) at $0.1<z<0.4$, $D_M/r_d$ and $D_H/r_d$ from two Luminous Red Galaxy (LRG) bins at $0.4<z<0.8$, a combined LRG+ELG tracer at $0.8<z<1.1$, ELG measurements at $1.1<z<1.6$, Quasar (QSO) clustering BAO at $0.8<z<2.1$, and Ly$\alpha$ forest BAO at $1.8<z<4.2$.
\begin{table}[h!]
    \centering
    \resizebox{\columnwidth}{!}{%
    \begin{tabular}{lccccc}
    \hline
    Tracer & $z_{\mathrm{eff}}$ & $D_{\mathrm{V}}/r_{\mathrm{d}}$ & $D_{\mathrm{M}}/r_{\mathrm{d}}$ & $D_{\mathrm{H}}/r_{\mathrm{d}}$ & $r_{\mathrm{M,H}}$ \\
    \hline
    BGS        & 0.295 & $7.942 \pm 0.075$ & --- & --- & --- \\
    LRG1       & 0.510 & --- & $13.588 \pm 0.167$ & $21.863 \pm 0.425$ & $-0.459$ \\
    LRG2       & 0.706 & --- & $17.351 \pm 0.177$ & $19.455 \pm 0.330$ & $-0.404$ \\
    LRG3+ELG1  & 0.934 & --- & $21.576 \pm 0.152$ & $17.641 \pm 0.193$ & $-0.416$ \\
    ELG2       & 1.321 & --- & $27.601 \pm 0.318$ & $14.176 \pm 0.221$ & $-0.434$ \\
    QSO        & 1.484 & --- & $30.512 \pm 0.760$ & $12.817 \pm 0.516$ & $-0.500$ \\
    Lya        & 2.330 & --- & $38.988 \pm 0.531$ & $8.632 \pm 0.101$ & $-0.431$ \\
    \hline
    \end{tabular}
    }
    \caption{BAO measurements from DESI BAO (DR2) used in this work \cite{DESI:2025zgx}.}
    \label{Tab:DR2_data}
\end{table}

\subsection{Cosmic Microwave Background (CMB)}
The CMB radiation provides a snapshot of the early Universe at the epoch of photon decoupling and serves as a key probe for testing cosmological models. We adopt the CMB distance priors from \textit{Planck} 2018 \cite{Wang:2007mza}, characterized by three parameters: the acoustic scale $l_a \equiv \pi r(z_\star)/r_s(z_\star)$, the shift parameter $R \equiv \sqrt{\Omega_m H_0^2}, r(z_\star)/c$, and the physical baryon density $\omega_b = \Omega_b h^2$ ($h \equiv H_0 / 100~\mathrm{km,s^{-1},Mpc^{-1}}$). Here, $l_a$ determines the peak spacing, while $R$ primarily affects their overall amplitude. The CMB information is summarized by the data vector $\mathbf{v} = (R,, l_a,, \omega_b)$ and its associated covariance matrix \cite{Zhai:2018vmm}. This compressed representation captures the essential geometric constraints from \textit{Planck} while minimizing prior-dependent effects.

\subsection{Supernova type Ia}
SNe~Ia serve as standardizable candles, providing precise measurements of the cosmic expansion history through the luminosity distance–redshift relation. In this work, we employ the PantheonPlus compilation \cite{Brout:2022vxf}, which comprises 1701 light curves of 1550 spectroscopically confirmed SNe~Ia from 18 surveys, spanning the redshift range $0.001 < z < 2.26$. This dataset is particularly valuable at low redshifts ($0.01 <z < 0.3$), where BAO constraints are limited by cosmic variance. Compared with the original Pantheon release, PantheonPlus provides improved photometric calibration, extended low-$z$ coverage, and a refined treatment of systematic uncertainties. In addition, we use the Union3 compilation, a comprehensive dataset of 2087 SNe~Ia from 24 samples covering $0.01 < z < 2.26$, standardized on a consistent distance scale using the SALT3 light-curve fitter. The 22 binned distance moduli of the Union3 sample are reported in \cite{Rubin:2023ovl}.

\section{Methodology} \label{sect:method}
To constrain the free parameters of the competing cosmological models and to do a rigorous determination of which model is most supported by observational data, we employ a comprehensive Bayesian statistical framework. Our methodology is divided into two main parts: parameter estimation and model comparison.

\subsection{Parameter Estimation and Likelihood Analysis}
The first part of our analysis is based on the determination of the distributions of posterior probability for the cosmological parameters of each model, given the observational data. For a given model with a set of parameters $\boldsymbol{\theta}$, the posterior probability is given by Bayes' theorem:
\begin{equation}
    P(\boldsymbol{\theta}|D) \propto P(D|\boldsymbol{\theta}) P(\boldsymbol{\theta}) = \mathcal{L}(D|\boldsymbol{\theta}) \pi(\boldsymbol{\theta})
\end{equation}
Here, $P(\boldsymbol{\theta}|D)$ is the posterior probability of the parameters, $\pi(\boldsymbol{\theta})$ is the prior probability distribution, encapsulating our initial knowledge of the parameters, and $\mathcal{L}(D|\boldsymbol{\theta})$ is the likelihood function.
Our total likelihood function, $\mathcal{L}_{\text{tot}}$, is obtained by multiplying the likelihoods of the different observational datasets, assuming they are independent:
\begin{equation}
    \mathcal{L}_{\text{tot}} = \mathcal{L}_{\text{DESI BAO}} \times \mathcal{L}_{\text{CMB}} \times \mathcal{L}_{\text{SN Ia}} ,
\end{equation}
where $\mathcal{L}_{\text{SN Ia}}$ refers to the likelihood function of the sample PantheonPlus or Union3.\\
To explore the multidimensional parameter space and map the posterior distributions, we employ a Markov Chain Monte Carlo (MCMC) sampling algorithm implemented using the publicly available \texttt{emcee} package \cite{Foreman_Mackey_2013}. From the obtained posterior distributions, we derive the best-fit values (e.g., the mean of each distribution) along with the corresponding confidence intervals at the $1\sigma$ ($68\%$), $2\sigma$ ($95\%$), and $3\sigma$ ($99.7\%$) levels. The analysis and visualization of the resulting chains are performed with the \texttt{GetDist} package \cite{Lewis:2019xzd}.

\subsection{Bayesian Model Comparison}
While obtaining the best-fit values of the cosmological parameters is crucial, a more profound question is which underlying cosmological model provides the best explanation for the cosmological data, especially when cosmological models have different numbers of parameters and differ in complexity. We note that a simple comparison based on the MLE (or minimum values $\chi^2$) is insufficient, as cosmological models with more free parameters often yield a better fit to the data, even if they lack physical justification (a phenomenon known as overfitting).
Although the Akaike Information Criterion (AIC) \cite{1100705} suggests an improvement by adding a penalty (twice the number of free parameters), it can be known as a simple approximation for model complexity.
As another method to overcome these limitations, we adopt the more robust and philosophically sound approach, namely the BE framework for model comparison.\\
The BE (also known as the marginal likelihood $\mathcal{Z}$) is defined as the probability of the data $D$ given the model $M$, averaged over all possible values of the parameters as follows:
\begin{equation}
    \mathcal{Z} = P(D|M) = \int \mathcal{L}(D|\boldsymbol{\theta}, M) \pi(\boldsymbol{\theta}|M) d\boldsymbol{\theta}
\end{equation}
The BE naturally considers Occam's razor. In general, we expect to see that a simple model makes very specific predictions, while a complex model spreads its predictions over a wide range of possibilities. Occam's razor says that a simpler model has high evidence if the observational data match its narrow predictions. While a complex model is penalized for wasting its predictive power on all the other possibilities that data did not support. Therefore, more complex models are penalized naturally unless for an extra complexity which is truly required by the data.  
To compare two cosmological models $M_0$ and $M_1$, we compute the ratio of their evidences, known as the Bayes Factor (BF):
\begin{equation}\label{eq:BF1}
    \text{BF}_{01} = \frac{\mathcal{Z}_0}{\mathcal{Z}_1}
\end{equation}
The BF value indicates a degree of support by observational data for one model over another model. We can interpret the power of this support by using Jeffreys' scale \cite{Jeffreys:1939xee}, where $\text{BF}>1$ ($\text{BF}<1$) indicates a preference for the model $M_0$ ($M_1$). The BF value allows us to make a quantitative statement about which cosmological model is most favored by the observational data. See Tables \ref{Tab:jeffrey1}, \ref{Tab:jeffrey2} for different statements based on the BF values in the context of Jeffreys' scale.
\begin{table}[h!]
    \centering
    \begin{tabular}{ccc}
    \hline
    $B_{0,1}$ & $\log_{10}(B_{0,1})$ & Strength of evidence \\
    \hline
    1 to 3.2   & 0 to 0.5   & Barely worth mentioning \\
    3.2 to 10  & 0.5 to 1.0 & Substantial \\
    10 to 32 & 1.0 to 1.5 & Strong \\
    32 to 100 & 1.5 to 2.0 & Very strong \\
    $> 100$     & $> 2.0$    & Decisive \\
    \hline
    \end{tabular}
    \caption{Jeffreys' scale for the strength of evidence in favor of simple model $M_0$ ($B_{0,1} > 1$).}
    \label{Tab:jeffrey1}
\end{table}

\begin{table}[h!]
    \centering
    \begin{tabular}{ccc}
    \hline
    $B_{0,1}$ & $\log_{10}(B_{0,1})$ & Strength of evidence \\
    \hline
    1 to 0.32   & 0 to -0.5   & Barely worth mentioning \\
    0.32 to 0.1 & -0.5 to -1.0 & Substantial \\
    0.1 to 0.032 & -1.0 to -1.5 & Strong \\
    0.032 to 0.01 & -1.5 to -2.0 & Very strong \\
    $< 0.01$     & $< -2.0$    & Decisive \\
    \hline
    \end{tabular}
    \caption{Jeffreys' scale for the strength of evidence in favor of complex model $M_1$ ($B_{0,1} < 1$).}
    \label{Tab:jeffrey2}
\end{table}
As mentioned above, in this work, extensions of the CPL parameterization are supposed to be studied from the statistical point of view. Overall, we have the following cases in our comparative analysis: standard $\Lambda$CDM cosmology, $w$CDM model, and CPL, CPL$^+$, CPL$^{++}$, Quadratic, and Cubic parameterizations for the equation of state of DE. 
In our comparative analysis, we consider a set of parameterizations referred to as nested models, meaning that one model (or parameterization) can be obtained from another by taking a specific limit or fixing certain parameters to constant values. For instance, the $\Lambda$CDM model ($w_{\Lambda}=-1$) is nested within the $w$CDM model ($w=\text{constant}$) by setting $w=-1$. Similarly, the CPL parametrization ($w_{\mathrm{de}}(a) = w_0 + w_a(1-a)$) reduces to $w$CDM when $w_a=0$. Extending this form, the CPL$^+$ parametrization ($w(a) = w_0 + w_a(1-a) + w_b(1-a)^2$) is nested within CPL$^{++}$ by setting $w_c=0$, while the standard CPL model is nested within CPL$^+$ by setting $w_b=0$. Analogously, the Quadratic and Cubic parameterizations can be viewed as nested extensions of the $w$CDM model. The hierarchical relationship in nested models is useful for understanding how increasing model complexity affects our confidence in the final assessment.\\
In the following section, we present our numerical results. We will report the best-fit values of the cosmological parameters obtained via MLE in an MCMC chain, followed by our primary statistical comparison of the models using the BE approach. To provide a supplementary check, we also report the AIC values, noting that agreement between the BF and AIC criteria strengthens the robustness of our conclusions.
\begin{table*}[ht]
    \centering
    \renewcommand{\arraystretch}{1.3}
    \begin{tabular}{lccccccc}
        \hline
        Model & $\Omega_b$ & $\Omega_{DM}$ & $h$ & $w_0$ & $w_a$ & $w_b$ & $w_c$ \\
        \hline
        $\Lambda$CDM 
        & [0, 0.1] & [0, 1] & [0, 2] & -- & -- & -- & -- \\
        $w$CDM 
        & [0, 0.1] & [0, 1] & [0, 2] & [-5, 5] & -- & -- & -- \\
        CPL 
        & [0, 0.1] & [0, 1] & [0, 2] & [-5, 5] & [-10, 10] & -- & -- \\
        Quadratic
        & [0, 0.1] & [0, 1] & [0, 2] & [-5, 5] & -- & [-30, 30] & -- \\
        Cubic 
        & [0, 0.1] & [0, 1] & [0, 2] & [-5, 5] & -- & -- & [-50, 50] \\
        CPL$^+$
        & [0, 0.1] & [0, 1] & [0, 2] & [-5, 5] & [-10, 10] & [-30, 30] & -- \\
        CPL$^{++}$ 
        & [0, 0.1] & [0, 1] & [0, 2] & [-5, 5] & [-10, 10] & [-30, 30] & [-50, 50] \\
        \hline
    \end{tabular}
    \caption{Priors adopted for the free parameters of each cosmological model considered in this work.}
    \label{tab:priors}
\end{table*}

\begin{table*}[ht]
    \centering
    \renewcommand{\arraystretch}{1.5}
    \begin{tabular}{lccc}
        \hline
        Model & $\Omega_m$ & $H_0$ & $M$ \\
        \hline
        $w$CDM 
        & $0.282^{+0.004, +0.008, +0.013}_{-0.004, -0.008, -0.012}$
        & $69.9^{+0.5, +1.0, +1.6}_{-0.5, -1.0, -1.6}$
        & $-19.369^{+0.013, +0.025, +0.037}_{-0.013, -0.025, -0.038}$ \\
        CPL
        & $0.291^{+0.005, +0.010, +0.015}_{-0.005, -0.010, -0.014}$
        & $69.3^{+0.6, +1.1, +1.7}_{-0.5, -1.1, -1.6}$
        & $-19.366^{+0.013, +0.026, +0.038}_{-0.013, -0.025, -0.038}$ \\
        Quadratic 
        & $0.291^{+0.005, +0.010, +0.015}_{-0.005, -0.009, -0.014}$
        & $69.3^{+0.5, +1.1, +1.7}_{-0.6, -1.1, -1.6}$
        & $-19.367^{+0.013, +0.026, +0.039}_{-0.013, -0.025, -0.038}$ \\
        Cubic 
        & $0.291^{+0.005, +0.010, +0.015}_{-0.005, -0.009, -0.014}$
        & $69.4^{+0.5, +1.1, +1.7}_{-0.5, -1.1, -1.6}$
        & $-19.368^{+0.013, +0.025, +0.039}_{-0.013, -0.025, -0.038}$ \\
        CPL$^+$
        & $0.290^{+0.005, +0.010, +0.015}_{-0.005, -0.010, -0.015}$
        & $69.6^{+0.6, +1.2, +1.8}_{-0.6, -1.1, -1.7}$
        & $-19.370^{+0.013, +0.026, +0.040}_{-0.013, -0.026, -0.039}$ \\
        CPL$^{++}$ 
        & $0.292^{+0.005, +0.010, +0.016}_{-0.005, -0.010, -0.015}$
        & $69.5^{+0.6, +1.2, +1.8}_{-0.6, -1.2, -1.7}$
        & $-19.368^{+0.013, +0.026, +0.039}_{-0.013, -0.025, -0.038}$ \\
        \hline
    \end{tabular}
    \caption{Observational constraints on $\Omega_m$, $H_0$ and nuisance parameter $M$ for different DE parameterizations, utilizing CMB + DESI BAO (DR2) +SN Ia PantheonPlus combination.}
    \label{Tab:cosmology_parameters_pan}
\end{table*}

\begin{table*}[ht]
    \centering
    \renewcommand{\arraystretch}{1.5}
    \resizebox{\textwidth}{!}{%
    \begin{tabular}{lcccc}
        \hline
        Model & $w_0$ & $w_a$ & $w_b$ & $w_c$ \\
        \hline
        $w$CDM 
        & $-0.995^{+0.020, +0.040, +0.061}_{-0.020, -0.040, -0.061}$ 
        & -- & -- & -- \\
        CPL
        & $-0.81^{+0.05, +0.10, +0.15}_{-0.05, -0.10, -0.15}$ 
        & $-0.73^{+0.19, +0.37, +0.55}_{-0.20, -0.40, -0.62}$ 
        & -- & -- \\
        Quadratic 
        & $-0.88^{+0.04, +0.07, +0.11}_{-0.03, -0.07, -0.10}$ 
        & -- 
        & $-1.4^{+0.4, +0.7, +0.9}_{-0.4, -0.8, -1.3}$ 
        & -- \\
        Cubic 
        & $-0.898^{+0.031, +0.062, +0.096}_{-0.030, -0.060, -0.090}$ 
        & -- & -- 
        & $-2.7^{+0.7, +1.4, +1.9}_{-0.9, -1.8, -2.9}$ \\
        CPL$^+$
        & $-1.06^{+0.11, +0.21, +0.32}_{-0.12, -0.24, -0.37}$ 
        & $1.8^{+1.2, +2.4, +3.8}_{-1.0, -2.0, -2.9}$ 
        & $-4.5^{+1.9, +3.5, +4.9}_{-2.2, -4.7, -7.6}$ 
        & -- \\
       CPL$^{++}$ 
        & $-0.86^{+0.17, +0.32, +0.45}_{-0.20, -0.39, -0.59}$ 
        & $-1.8^{+3.2, +6.2, +8.8}_{-2.7, -4.5, -5.7}$ 
        & $11^{+12, +18, +19}_{-13, -24, -32}$ 
        & $-17^{+14, +26, +33}_{-15, -24, -30}$ \\
        \hline
    \end{tabular}}
    \caption{Observational constraints on DE parameters $w_0, w_a, w_b, w_c$ for different DE parametrizations, utilizing the combination of CMB + DESI BAO (DR2) + SN Ia PantheonPlus dataset.}
    \label{Tab:DE_parameters_pan}
\end{table*}

\section {Numerical results}\label{Sect:Numerical_results}
In this section, we present our numerical results for the cosmological analysis, including the observational constraints on the parameters of different DE parametrizations, and perform a rigorous model comparison. We utilize two dataset combinations (i). CMB+DESI BAO (DR2)+SN Ia PantheonPlus and (ii). CMB+DESI BAO (DR2)+SN Ia Union3. For all parametrizations under study, we adopt the uniform priors for the free cosmological parameters as shown in Table \ref{tab:priors}. Furthermore, for PantheonPlus data, the absolute magnitude ($M$) is treated as a nuisance parameter, with a uniform prior of $M \in [-20, -18]$. The widths of the priors are carefully chosen to be broad enough to prevent artificial truncation of the posterior distributions, while at the same time narrow enough to avoid imposing undue penalties on the extended CPL parametrizations. This balanced choice ensures that the observational data can effectively constrain the free parameters.

\subsection{CMB + DESI BAO (DR2) + PantheonPlus combination}
We first report the observational constraints on the cosmological parameters and then perform the statistical comparison between various DE parametrizations under consideration. 

\subsubsection{Constraints to parameters}
We present the best-fit values and 1-3$\sigma$ confidence intervals for the cosmological parameters ($\Omega_m$, $H_0$, and $M$) in Table \ref{Tab:cosmology_parameters_pan}. Our constraints are generally consistent across all DE parameterizations, indicating a robust determination of cosmology in the background level. In addition, the constraints on the parameters related to the equation of state of DE ($w_0$, $w_a$, $w_b$, and $w_c$) for each case are summarized in Table \ref{Tab:DE_parameters_pan}.\\
The standard CPL parametrization shows a best-fit of $w_0 = -0.81$ and $w_a = -0.73$. Interestingly, the $\Lambda$CDM point ($w_0=-1, w_a=0$) lies well outside the 3$\sigma$ confidence region for this parametrization, showing a statistical preference for dynamically evolving DE. (see also \cite{DESI:2024mwx,DESI:2025zgx}). Notice that the statistical deviation of the maximum likelihood in full space parameters from the $\Lambda$CDM model supports this preference.\\
For the extended parameterizations of CPL, like CPL$^+$ and CPL$^{++}$, the additional parameters ($w_b$ and $w_c$) are introduced. We observe that the central best-fit values deviate from zero, while their uncertainties are substantial. Moreover, we see the large uncertainties for higher-order parameters ($w_b$ and $w_c$). The large uncertainties in the DE parameterizations for CPL$^+$ and CPL$^{++}$ compared to the CPL case are expected, as both cases include more free parameters that we aim to constrain using the same datasets.\\
We emphasize that merely observing a deviation from the $\Lambda$CDM point ($-1, 0$) in the $w_0 -w_a$ plane does not necessarily indicate the possibility of an evolving DE paradigm. In the extended CPL$^+$ and CPL$^{++}$ parameterizations, the inclusion of additional parameters $w_b$ and $w_c$ introduces significantly larger uncertainties and degeneracies. Consequently, the tightness of observational constraints in the $w_0 -w_a$ plane becomes substantially weaker, with confidence regions broad enough to encompass the $\Lambda$CDM point. We need a comprehensive comparison across the full parameter space, employing statistical indicators such as the MLE or the BE approach (see the following sections), to address the possibility of potential deviations from the standard $\Lambda$CDM cosmology.\\
The alternative two-parameter scenarios, Quadratic ($w_0, w_b$) and Cubic ($w_0, w_c$) parametrizations, also show a preference for $w_0 > -1$  and $w_b<0$ (Quadratic) $w_c<0$ (Cubic) with more than $3\sigma$ from the fixed point $(-1.0, 0.0)$, consistent with the findings of CPL.
\begin{table}[ht]
	\centering
	\renewcommand{\arraystretch}{1.3}
	\begin{tabular}{lcccc}
		\hline
		Model & $\chi^2_{Best}$ & $\Delta \chi^2_{\text{Best}}$ & $\Delta \text{AIC}$ & $\text{deviation}$ \\
		\hline
		$\Lambda \text{CDM}$ 
		& $1582.89$ & -- & -- & --\\
		$\text{wCDM}$ 
		& $1582.83$ & $-0.06$ & $1.94$ & $-0.9 \sigma$\\
		$\text{CPL}$ 
		& $1573.88$ & $-9.01$ & $-5.01$ & $2.4 \sigma$\\
		$\text{Quadratic}$ 
		& $1571.66$ & $-11.23$ & $-7.23$ & $2.7 \sigma$\\
		$\text{Cubic}$ 
		& $1569.85$ & $-13.04$ & $-9.04$ & $3.0 \sigma$\\
		$\text{CPL}^+$ 
		& $1569.29$ & $-13.60$ & $-7.60$ & $2.7 \sigma$\\
		$\text{CPL}^{++}$ 
		& $1568.94$ & $-13.95$ & $-5.95$ & $2.5 \sigma$\\
		\hline
	\end{tabular}
	\caption{The values of $\chi_{\text{Best}}^2$, $\Delta \chi_{\text{Best}}^2$, $\Delta \text{AIC}$ and deviation in different parametrizations, computed using the CMB + DESI BAO (DR2) + PantheonPlus dataset.}
	\label{Tab:x2_pan}
\end{table}

\begin{table}[ht]
	\centering
	\renewcommand{\arraystretch}{1.3}
	\begin{tabular}{lcc}
		\hline
		Model (0) vs Model (1) & $\text{BF}$ & $\log_{10}(\text{BF})$ \\
		\hline
		$\Lambda \text{CDM}$ vs $\text{wCDM}$ 
		& $139$ & $2.14$  \\
		$\text{wCDM}$ vs $\text{CPL}$  
		& $0.022$ & $-1.66$  \\
            $\text{wCDM}$ vs $\text{Quadratic}$  
		& $0.054$ & $-1.27$  \\
            $\text{wCDM}$ vs $\text{Cubic}$ 
		& $0.115$ & $-0.94$  \\
            $\text{CPL}$ vs $\text{CPL}^+$
		& $0.824$ & $-0.08$  \\
            $\text{CPL}$ vs $\text{CPL}^{++}$ 
		& $6.153$ & $0.79$ \\
        \hline
           $\Lambda \text{CDM}$  vs $\text{CPL}$
		& $2.307$ & $0.36$  \\
           $\Lambda \text{CDM}$  vs $\text{Quadratic}$ 
		& $4.251$ & $0.63$  \\
          $\Lambda \text{CDM}$ vs  $\text{Cubic}$ 
		& $6.024$ & $0.78$  \\
		\hline
	\end{tabular}
    \caption{Bayes factors (BF) and log$_{10}$(BF), computed using the CMB + DESI BAO (DR2) + PantheonPlus dataset.}
    \label{Tab:BF_pan}
\end{table}

\subsubsection{Model Comparison:  MLE and AIC Analysis}\label{Sub:subAIC}
We begin by performing a model comparison based on the maximum likelihood estimation (MLE) and the Akaike Information Criterion (AIC). The results are summarized in Table~\ref{Tab:x2_pan}. In this context, $\Delta \chi^2_{\text{Best}}$ is defined as $
\Delta \chi^2_{\text{Best}} = \chi^2_{\text{Best}}(\text{model}) - \chi^2_{\text{Best}}(\Lambda\text{CDM}),
$ while $\Delta \text{AIC}$ denotes 
$\Delta \text{AIC} = \text{AIC}(\text{model}) - \text{AIC}(\Lambda\text{CDM})$. Assuming Gaussian errors, the likelihood is related to the chi-squared statistic via
$\chi^2 \propto -2 \ln \mathcal{L}$, so maximizing the likelihood is equivalent to minimizing $\chi^2$.
For the baseline $\Lambda$CDM cosmology, we obtain $\chi^2_{\text{Best}} = 1582.89$. The wCDM extension yields only a negligible improvement with $\Delta \chi^2 = -0.06$, but is penalized by the AIC, giving $\Delta \text{AIC} = +1.94$; thus, it shows no significant preference over $\Lambda$CDM. In contrast, the CPL parameterization achieves $\chi^2_{\text{Best}} = 1573.88$, corresponding to an improvement of $\Delta \chi^2 = -9.01$ relative to $\Lambda$CDM. This translates into $\Delta \text{AIC} = -5.01$, indicating a moderate statistical preference for the CPL model, with a significance level of approximately $2.4\sigma$ compared to the standard $\Lambda$CDM (see also \cite{DESI:2024mwx}).\\
All more complex parametrizations, including CPL$^+$, CPL$^{++}$, Quadratic, and Cubic, yield a better fit to the observational data than the concordance $\Lambda$CDM cosmology, with $\Delta\chi^2_{\text{Best}}$ values ranging from $-11.23$ to $-13.95$. Since all these parametrizations are generalizations of the CPL parametrization, it is instructive to compare them directly with CPL. In particular, the Quadratic model yields $\Delta \text{AIC} = \text{AIC} - \text{AIC}_{\Lambda \text{CDM}} = -7.23$, and the Cubic model gives $\Delta \text{AIC} = \text{AIC} - \text{AIC}_{\Lambda \text{CDM}} = -9.04$, both achieving lower AIC values than standard CPL. This suggests that the observational data favor these functional forms for the DE equation of state. In these cases, we observe deviations of $2.7\sigma$ and $3.0\sigma$ from the $\Lambda$CDM cosmology, supporting the result obtained using the CPL parametrization. For CPL$^+$ and CPL$^{++}$, the fit to the data remains unchanged despite the inclusion of one and two additional parameters, respectively, compared to CPL. Therefore, the evidence supporting the need for increased complexity in the equation of state beyond two parameters appears less compelling from the viewpoint of MLE.

\subsubsection{Model Comparison: Bayesian Evidence Analysis}\label{sec:BF}
We now perform a detailed model comparison using the BE, with the Bayes Factors (BF) and their logarithmic values presented in Table~\ref{Tab:BF_pan}. The strength of the BF is interpreted using Jeffreys' scale.\\
First, we compare the one-parameter extension, the wCDM model, against the baseline $\Lambda$CDM cosmology. The BE analysis for $\Lambda$CDM against wCDM yields $\log_{10}(\text{BF}) = 2.14$, indicating \textit{Decisive} evidence in favor of the $\Lambda$CDM model.\newline
Next, we investigate the two-parameter parametrizations (CPL, Quadratic, and Cubic) against the simpler one-parameter wCDM model, as they can be seen as generalizations of it. The BF of wCDM against CPL is $\log_{10}(\text{BF}) = -1.66$, which indicates \textit{Decisive} evidence against the wCDM model, meaning the CPL parametrization is strongly preferred. Similarly, for wCDM against Quadratic, the result $\log_{10}(\text{BF}) = -1.27$ provides \textit{Strong} evidence against the wCDM model in favor of the Quadratic parametrization. The wCDM versus Cubic comparison, with $\log_{10}(\text{BF}) = -0.94$, shows \textit{Substantial} evidence against the wCDM model. Hence, in agreement with our results based on the MLE, we observe that the increased complexity and larger parameter volume of these two-parameter parametrizations are not heavily penalized by the BE, and all of them are favored compared to the simpler wCDM model.\\
We now examine whether extending the standard CPL parameterization with higher-order terms is supported by the observational data. The comparison of CPL against CPL$^+$ yields $\log_{10}(\text{BF}) = -0.08$, indicating \textit{Barely worth mentioning} on Jeffreys' scale, and implying that there is no significant preference between these two parametrizations. Hence, including of the quadratic term ($w_b$) to the CPL case is not observationally supported.\\
The Bayesian comparison between baseline CPL and CPL$^{++}$ gives $\log_{10}(\text{BF}) = 0.79$, representing \textit{Substantial} evidence favoring CPL (simple model) over CPL$^{++}$ (complex model). Therefore, the added quadratic and cubic terms are not essentially supported by observational data.\\
Finally, it is worthwhile to compare the two-parameter $w_0w_a$ parameterizations with the concordance $\Lambda$CDM cosmology. Notice that in our Bayesian statistics, the two-parameter $w_0w_a$ parameterizations were favored over both the single-parameter wCDM and the more complex CPL$^+$ and CPL$^{++}$ parametrizations. Our results yield $\log_{10}(\text{BF}) = 0.36, 0.63 \; \& \; 0.78$, respectively, for concordance $\Lambda$CDM cosmology against the CPL, Quadratic, and Cubic parametrizations. Based on the strength of Jeffreys' scale, there is no strong support for choosing the $w_0w_a$ parameterizations over the standard $\Lambda$CDM cosmology. This result is in agreement with our conclusion based on the MLE analysis in the previous section, where we could not observe a strong deviation (more than $3\sigma$) from the $\Lambda$CDM cosmology.\\
  
\subsection{Analysis with CMB + DESI BAO (DR2) + Union3 Data}
In this subsection, we replace the PantheonPlus SN~Ia sample with the Union~3 compilation and repeat our analysis. The same cosmological priors listed in Table~\ref{tab:priors} are adopted. We then compare the results with those obtained from the PantheonPlus compilation in the previous subsection, in order to assess the robustness of our findings.

\subsubsection{Parameter Constraints}
The best-fit values for $\Omega_m$ and $H_0$ are reported in Table~\ref{Tab:cosmology_parameters_union}, while the DE parameter constraints are summarized in Table~\ref{Tab:DE_parameters_union}. The constraints on $\Omega_m$ and $H_0$ are broadly consistent with those obtained from the PantheonPlus sample, indicating a stable background cosmology across both SN~Ia compilations. In contrast, the constraints on the DE parameters reveal some notable differences. For the CPL parametrization, the best-fit values are $w_0 = -0.63$ and $w_a = -1.13$, suggesting that the Union~3 sample favors a stronger evolution and a more pronounced phantom crossing compared to the PantheonPlus data. In the cases of CPL$^{+}$ and CPL$^{++}$ parametrizations, which introduce higher-order parameters, the uncertainties in the DE constraints become substantial. This indicates that the current combination of observational data is insufficient to place tight bounds on the additional degrees of freedom.

\subsubsection{Model Comparison: MLE and AIC Analysis}
Our numerical results for the model comparison based on the MLE and AIC criteria are reported in Table~\ref{Tab:x2_union}. In general, our findings support a stronger preference for evolving DE, as detailed below. For the wCDM model, we observe no significant difference (deviation less than $1\sigma$) from the $\Lambda$CDM cosmology, consistent with our prediction using the PantheonPlus sample in the data combination discussed in the previous section. In the case of the CPL parametrization, the preference for an evolving DE scenario is more pronounced with the CMB + DESI BAO (DR2) + Union~3 combination. Statistically, the CPL parametrization achieves a remarkable improvement of $\Delta\chi^2 = -17.16$ over the $\Lambda$CDM model, corresponding to $\Delta\text{AIC} = -13.16$. This represents a statistical preference of $3.6\sigma$, which is significantly stronger than the $2.4\sigma$ preference found with the PantheonPlus sample. We obtain similar results for other DE parametrizations, including the Quadratic and Cubic cases with two degrees of freedom, and CPL$^+$ and CPL$^{++}$ cases with three and four degrees of freedom, respectively. We emphasize that the more complex CPL$^+$ ($\Delta\text{AIC} = -13.27$) and CPL$^{++}$ ($\Delta\text{AIC} = -11.56$) parametrizations do not provide better AIC values than the simpler CPL, Quadratic, and Cubic parametrizations. This indicates that there is no statistical support for extending the CPL functional form with additional higher-order terms. In other words, the presence of extra free parameters in these parametrizations does not play a significant role in achieving a better fit to the data. Consequently, the level of statistically significant deviation from the $\Lambda$CDM cosmology for CPL$^+$ and CPL$^{++}$ is not greater than that of the standard CPL parametrization.\\
Another point is that the best-fit values of the additional parameters $w_b$ and $w_c$ in the CPL$^+$ and CPL$^{++}$ parametrizations are consistent with $w_b = w_c = 0$ at the $2\sigma$ level. This indicates that introducing these extra degrees of freedom into the $w_0w_a$ parametrization does not have a significant impact, since at the $2\sigma$ level no meaningful deviation is observed (for a visual representation of the DE parameters, we refer the reader to Appendix~\ref{App:appendix}).
\begin{table*}[ht]
    \centering
    \renewcommand{\arraystretch}{1.5}
    \begin{tabular}{lcc}
        \hline
        Model & $\Omega_m$ & $H_0$ \\
        \hline
        $w$CDM 
        & $0.288^{+0.005, +0.010, +0.016}_{-0.005, -0.010, -0.015}$ 
        & $69.2^{+0.7, +1.3, +2.0}_{-0.7, -1.3, -1.9}$ \\
        $\text{CPL}$ 
        & $0.313^{+0.008, +0.016, +0.024}_{-0.008, -0.016, -0.023}$ 
        & $66.8^{+0.8, +1.7, +2.5}_{-0.8, -1.6, -2.4}$ \\
        $\text{Quadratic}$ 
        & $0.310^{+0.008, +0.015, +0.023}_{-0.008, -0.014, -0.021}$ 
        & $67.2^{+0.8, +1.6, +2.3}_{-0.8, -1.5, -2.2}$ \\
        $\text{Cubic}$ 
        & $0.308^{+0.007, +0.014, +0.022}_{-0.007, -0.014, -0.021}$ 
        & $67.4^{+0.8, +1.5, +2.3}_{-0.7, -1.4, -2.1}$ \\
        $\text{CPL}^+$ 
        & $0.307^{+0.009, +0.018, +0.027}_{-0.009, -0.017, -0.026}$ 
        & $67.6^{+1.0, +2.0, +3.0}_{-1.0, -1.9, -2.8}$ \\
        $\text{CPL}^{++}$ 
        & $0.311^{+0.010, +0.019, +0.029}_{-0.010, -0.019, -0.029}$ 
        & $67.2^{+1.1, +2.2, +3.4}_{-1.0, -2.0, -2.9}$ \\
        \hline
    \end{tabular}
    \caption{Same as Table \ref{Tab:cosmology_parameters_pan}, But for CMB + DESI BAO (DR2) + Union3 dataset.}
    \label{Tab:cosmology_parameters_union}
\end{table*}

\begin{table*}[ht]
    \centering
    \renewcommand{\arraystretch}{1.5}
    \resizebox{\textwidth}{!}{%
    \begin{tabular}{lcccc}
        \hline
        Model & $w_0$ & $w_a$ & $w_b$ & $w_c$ \\
        \hline
        $w$CDM 
        & $-0.973^{+0.025, +0.048, +0.072}_{-0.025, -0.050, -0.077}$ 
        & -- & -- & -- \\
        $\text{CPL}$ 
        & $-0.63^{+0.08, +0.16, +0.25}_{-0.08, -0.16, -0.24}$ 
        & $-1.13^{+0.26, +0.50, +0.74}_{-0.27, -0.56, -0.86}$ 
        & -- & -- \\
        $\text{Quadratic}$ 
        & $-0.76^{+0.05, +0.11, +0.16}_{-0.05, -0.10, -0.15}$ 
        & -- 
        & $-1.9^{+0.5, +0.8, +1.2}_{-0.5, -1.0, -1.6}$ 
        & -- \\
        $\text{Cubic}$ 
        & $-0.81^{+0.05, +0.09, +0.14}_{-0.05, -0.09, -0.13}$ 
        & -- & -- 
        & $-3.4^{+0.9, +1.6, +2.2}_{-1.0, -2.2, -3.5}$ \\
        $\text{CPL}^{+}$ 
        & $-0.88^{+0.18, +0.34, +0.52}_{-0.19, -0.38, -0.58}$ 
        & $1.0^{+1.5, +3.2, +4.9}_{-1.3, -2.5, -3.7}$ 
        & $-3.5^{+2.2, +4.0, +5.7}_{-2.6, -5.6, -8.9}$ 
        & -- \\
        $\text{CPL}^{++}$ 
        & $-0.64^{+0.27, +0.50, +0.69}_{-0.31, -0.61, -0.91}$ 
        & $-3^{+4, +8, +11}_{-3, -6, -7}$ 
        & $10^{+13, +19, +20}_{-13, -25, -34}$ 
        & $-14^{+13, +24, +31}_{-13, -22, -29}$ \\
        \hline
    \end{tabular}}
    \caption{Same as Table \ref{Tab:DE_parameters_pan}, But for CMB + DESI BAO (DR2) + Union3 dataset.}
    \label{Tab:DE_parameters_union}
\end{table*}

\begin{table}[ht]
	\centering
	\renewcommand{\arraystretch}{1.3}
	\begin{tabular}{lcccc}
		\hline
		Model & $\chi^2_{\text{Best}}$ & $\Delta \chi^2_{\text{Best}}$ & $\Delta \text{AIC}$ & $\text{deviation}$ \\
		\hline
		$\Lambda$CDM 
		& $78.36$ & -- & -- & --\\
		$\text{wCDM}$ 
		& $77.20$ & $-1.16$ & $0.84$ & $0.6 \sigma$\\
		$\text{CPL}$ 
		& $61.20$ & $-17.16$ & $-13.16$ & $3.6 \sigma$\\
		$\text{Quadratic}$ 
		& $58.71$ & $-19.65$ & $-15.65$ & $3.9 \sigma$\\
		$\text{Cubic}$ 
		& $58.88$ & $-19.48$ & $-15.48$ & $3.9 \sigma$\\
		$\text{CPL}^+$ 
		& $59.09$ & $-19.27$ & $-13.27$ & $3.5 \sigma$\\
		$\text{CPL}^{++}$ 
		& $58.80$ & $-19.56$ & $-11.56$ & $3.2 \sigma$\\
		\hline
	\end{tabular}
	\caption{Same as Table \ref{Tab:x2_pan}, But for CMB + DESI BAO (DR2) + Union3 dataset.}
	\label{Tab:x2_union}
\end{table}

\begin{table}[ht]
	\centering
	\renewcommand{\arraystretch}{1.3}
	\begin{tabular}{lcc}
		\hline
		Model (0) vs Model (1) & $\text{BF}$ & $\log_{10}(\text{BF})$ \\
		\hline
		  $\Lambda$CDM vs wCDM  
		& $63.3$ & $1.80$ \\
		wCDM vs CPL  
		& $0.003$ & $-2.52$ \\
            wCDM vs Quadratic 
		& $0.017$ & $-1.78$ \\
            wCDM vs Cubic 
		& $0.062$ & $-1.21$ \\
           CPL vs CPL$^+$  
		& $2.565$ & $0.41$ \\
           CPL vs CPL$^{++}$ 
		& $8.925$ & $0.95$ \\
        \hline
             $\Lambda$CDM  vs CPL 
		& $0.065$ & $-1.18$ \\
            $\Lambda$CDM  vs Quadratic  
		& $0.142$ & $-0.85$ \\
            $\Lambda$CDM  vs Cubic  
		& $0.333$ & $-0.48$ \\
		\hline
	\end{tabular}
    \caption{Same as Table \ref{Tab:BF_pan}, But for CMB + DESI BAO (DR2) + Union3 dataset.}
    \label{Tab:BF_union}
\end{table}

\subsubsection{Model Comparison: Bayesian Evidence Analysis}
Our numerical results based on the BE approach using the combination of CMB + DESI BAO (DR2) + Union~3 datasets are reported in Table~\ref{Tab:BF_union}. We observe that the BF value (ratio of BE of $\Lambda$CDM (simple model) to wCDM (complex model)) is $\log_{10}(\text{BF}) = 1.80$, which provides \textit{Very Strong} evidence in favor of the $\Lambda$CDM model over the wCDM model, in agreement with our Bayesian analysis using the CMB + DESI BAO (DR2) + PantheonPlus combination in the previous section. This result also supports our finding based on the MLE using the CMB + DESI BAO (DR2) + Union~3 datasets. The Bayesian comparison of wCDM (simple model) and CPL (complex model) yields $\log_{10}(\text{BF}) = -2.52$, providing \textit{Decisive} evidence against the wCDM model and strongly favoring the CPL parametrization as a complex model. Hence, the Occam's razor penalty is not large enough to reject the complex model. We obtain strong and very strong evidence against wCDM when utilizing the Cubic and Quadratic parametrizations as complex models in our Bayesian analysis, which supports the results obtained from the BE method using the PantheonPlus sample. Furthermore, the Bayesian comparison of CPL (as a simple model) with CPL$^+$ and CPL$^{++}$ (as complex models), respectively, gives $\log_{10}(\text{BF}) = 0.41$ and $\log_{10}(\text{BF}) = 0.95$, indicating that there is no compelling Bayesian support for adding higher-order terms to the CPL parameterization. Again, we observe consistency with our findings obtained using the PantheonPlus sample from the BE method in the previous section.
Finally, we compare the CPL, Quadratic, and Cubic parametrizations with the concordance $\Lambda$CDM cosmology, similar to the analysis performed in the previous section using the PantheonPlus data. Our numerical results show that there is \textit{Strong} ( $\log_{10}(\text{BF}) = -1.18$) and \textit{Substantial} ($\log_{10}(\text{BF}) = -0.85$) evidence, respectively, supporting the CPL and Quadratic parametrizations over the $\Lambda$CDM cosmology. In addition, in the case of comparison between $\Lambda$CDM cosmology and Cubic parameterization, we obtain $\log_{10}(\text{BF}) = -0.48$, indicating that there is no strong support behind the $\Lambda$CDM cosmology.

\section{Conclusion}\label{conlusion}
In this paper, we have carried out a comprehensive investigation into the necessity of DE parametrizations beyond two-parameter forms such as the CPL parametrization. We performed a detailed model comparison through a joint analysis of cosmological datasets, including Cosmic Microwave Background (CMB) distance priors, the second data release (DR2) of DESI Baryon Acoustic Oscillation (BAO) measurements, and two extensive Type~Ia Supernova (SN~Ia) compilations, PantheonPlus and Union~3. To assess the performance of different models, we employed two distinct yet complementary statistical frameworks: a likelihood-based approach using Maximum Likelihood Estimation (MLE) together with the Akaike Information Criterion (AIC), and a full Bayesian evidence (BE) calculation to rigorously account for model complexity and parametrization.\\
First and foremost, from the likelihood perspective, the AIC analysis shows a moderate to strong preference for the two-parameter CPL, Quadratic, and Cubic parametrizations over the $\Lambda$CDM cosmology, with the statistical significance of the deviation increasing up to $3.9\sigma$ using the Union3 sample of SN Ia. In addition, the BE supports this conclusion, favoring these evolving two-parameter DE parametrizations over the cosmological constant, $\Lambda$. The strength of the BE support ranges from \textit{Substantial} to \textit{Strong} depending on the specific parametrization and cosmological dataset, particularly with the Union3 data, where the CPL case is strongly preferred over the $\Lambda$CDM cosmology. Therefore, current cosmological data used in our analysis favor a dynamical description of DE over a standard cosmological constant, both from MLE and BE statistical methods.\\
The second major finding of our paper concerns the necessity of parameterizations beyond the standard two-parameter $w_0w_a$ dark energy framework. Using the MLE approach, we examined higher-order extensions of the CPL model (CPL$^+$ and CPL$^{++}$) and found that AIC analysis indicates the additional parameters are not justified by the marginal improvement in goodness-of-fit. Furthermore, the BE method provides even stronger support for this conclusion, favoring the simpler CPL parametrization over its more complex extensions. This outcome holds consistently across both SN~Ia datasets used in our analysis, suggesting that current cosmological observations do not warrant further complexity in modeling the DE equation of state.\\
We note that the PantheonPlus and Union3 SN~Ia catalogs are different in
calibration procedures, light-curve standardization, redshift coverage, and treatment of systematic uncertainties. These differences may lead to some variations in the inferred constraints on the parameters of DE, with the Union3 catalog showing a more significant preference for dynamical DE than
PantheonPlus sample. Analyzing both compilations provides a more robust and transparent assessment of whether current cosmological observations favor a deviation from the $\Lambda$CDM cosmology.\\
In summary, our joint analysis of the latest cosmological data indicates that the evidence for a dynamical DE component evolving with redshift is both significant and statistically consistent across different methodologies. However, our statistical results strongly suggest that DE evolution is well captured by the efficient two-parameter $w_0w_a$ parametrization. There is no statistical motivation to introduce higher-order terms or more complex forms beyond the CPL framework. Our findings, therefore, support recent indications of dynamical DE while simultaneously challenging the necessity of additional complexity beyond $w_0w_a$ parametrizations. We thus confirm that the well-chosen two-parameter $w_0w_a$ description remains the most effective and justified parametrization for studying the behavior of DE today.
\section{DATA AVAILABILITY}
The data used in this work are available publicly.
\section{acknowledgment}
MM would like to express gratitude to Ahmad Mehrabi for his valuable discussions.
The work of MM is based upon research funded by the Iran National Science
Foundation (INSF) under project No. 4041693. ZD is supported by the Korea Institute for Advanced Study (KIAS) under grant no.6G097301

\bibliographystyle{apsrev4-1}
\bibliography{ref}
\label{lastpage}

\section{Appendix: Confidence Contours for Dark Energy Parameters}
\label{App:appendix}
\begin{figure*}
    \centering
    \includegraphics[width=0.5\textwidth]{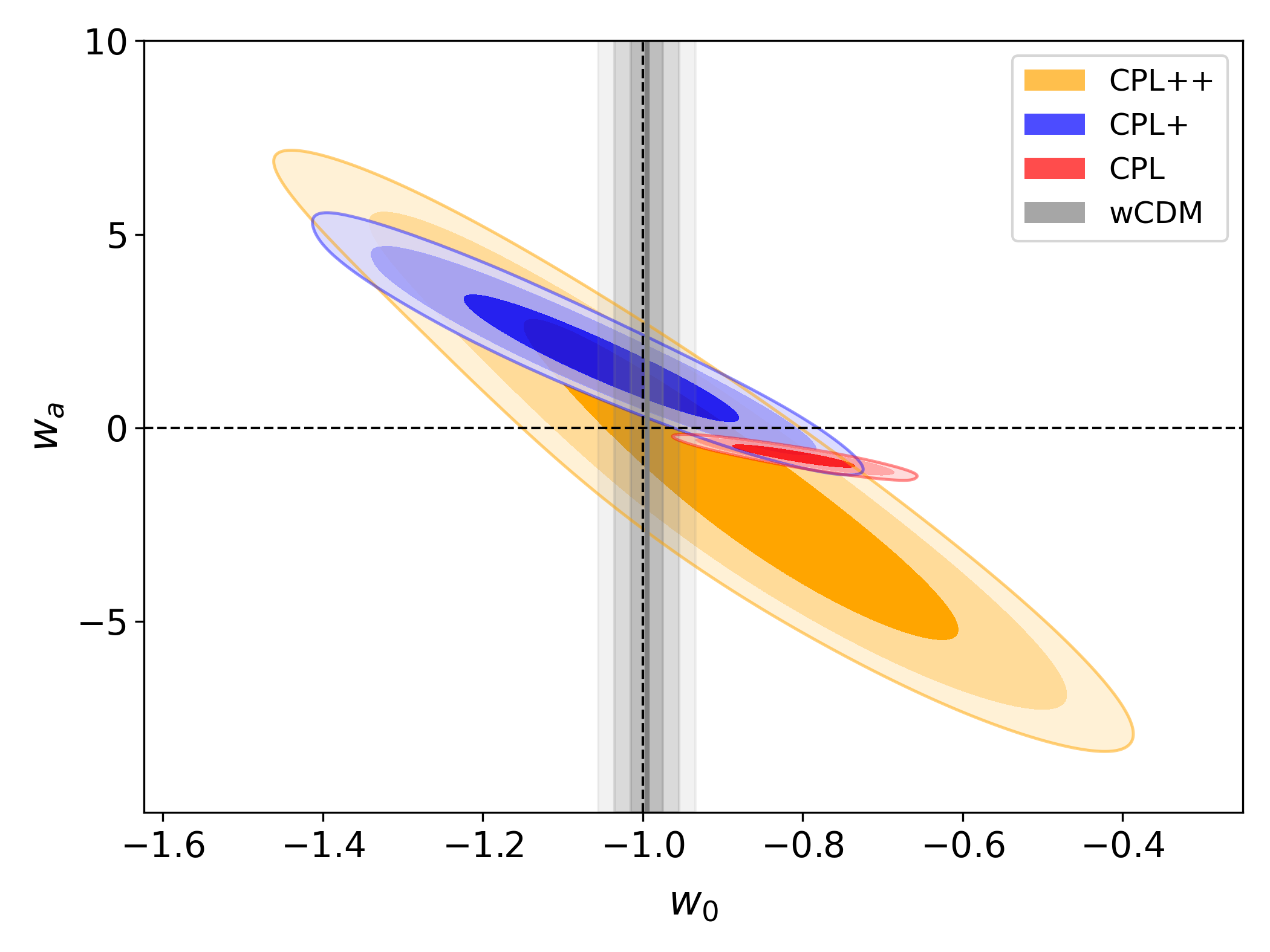}\includegraphics[width=0.5\textwidth]{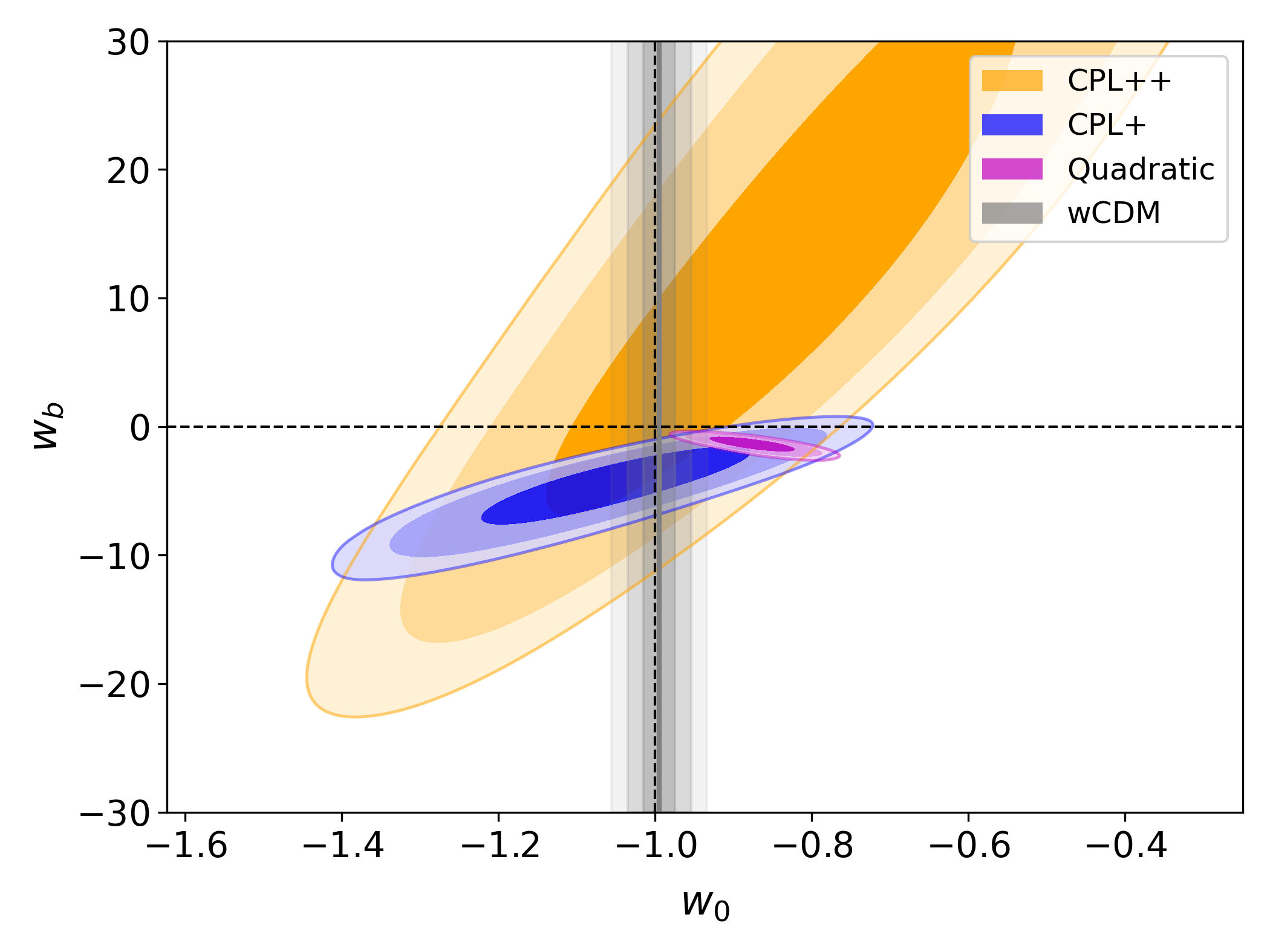}
    \includegraphics[width=0.5\textwidth]{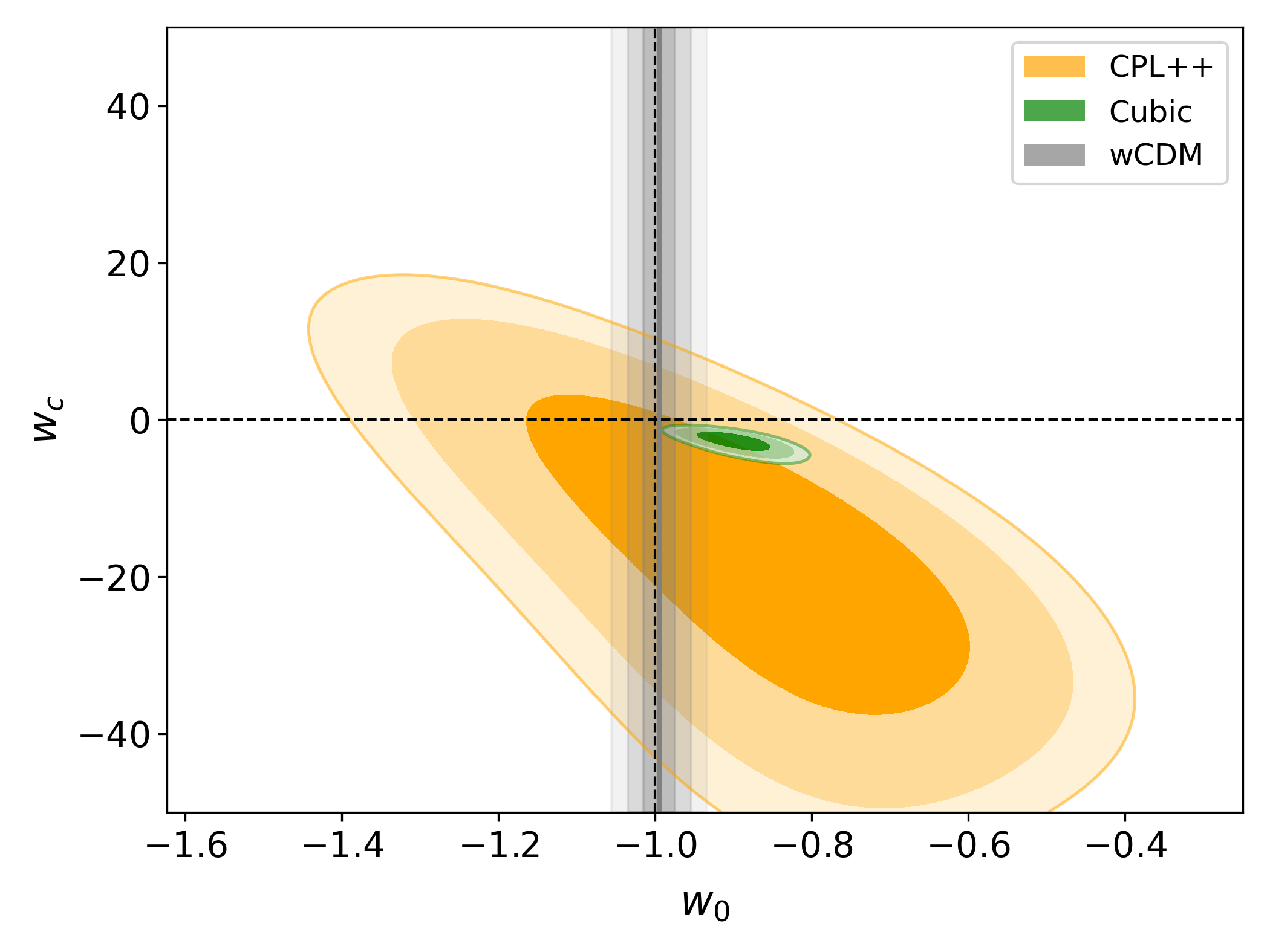}
    \caption{Two-dimensional marginalized posteriors within 1–2–3$\sigma$ confidence levels for different DE parameters ($w_0$, $w_a$, $w_b$, $w_c$) from CMB + DESI BAO (DR2) + PantheonPlus combination.}
    \label{fig:corner_pan}
\end{figure*}

\begin{figure*}
    \centering
    \includegraphics[width=0.5\textwidth]{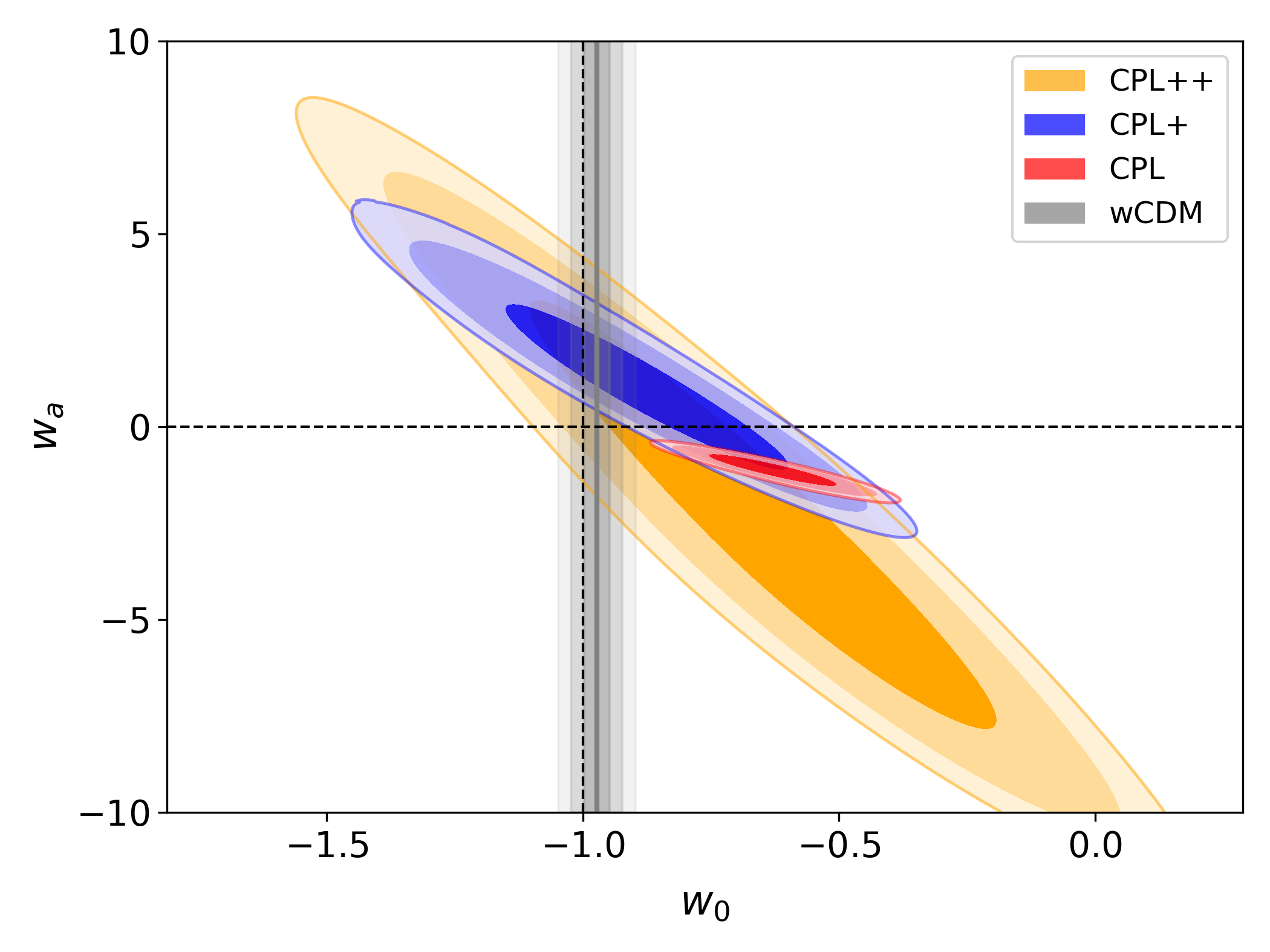}\includegraphics[width=0.5\textwidth]{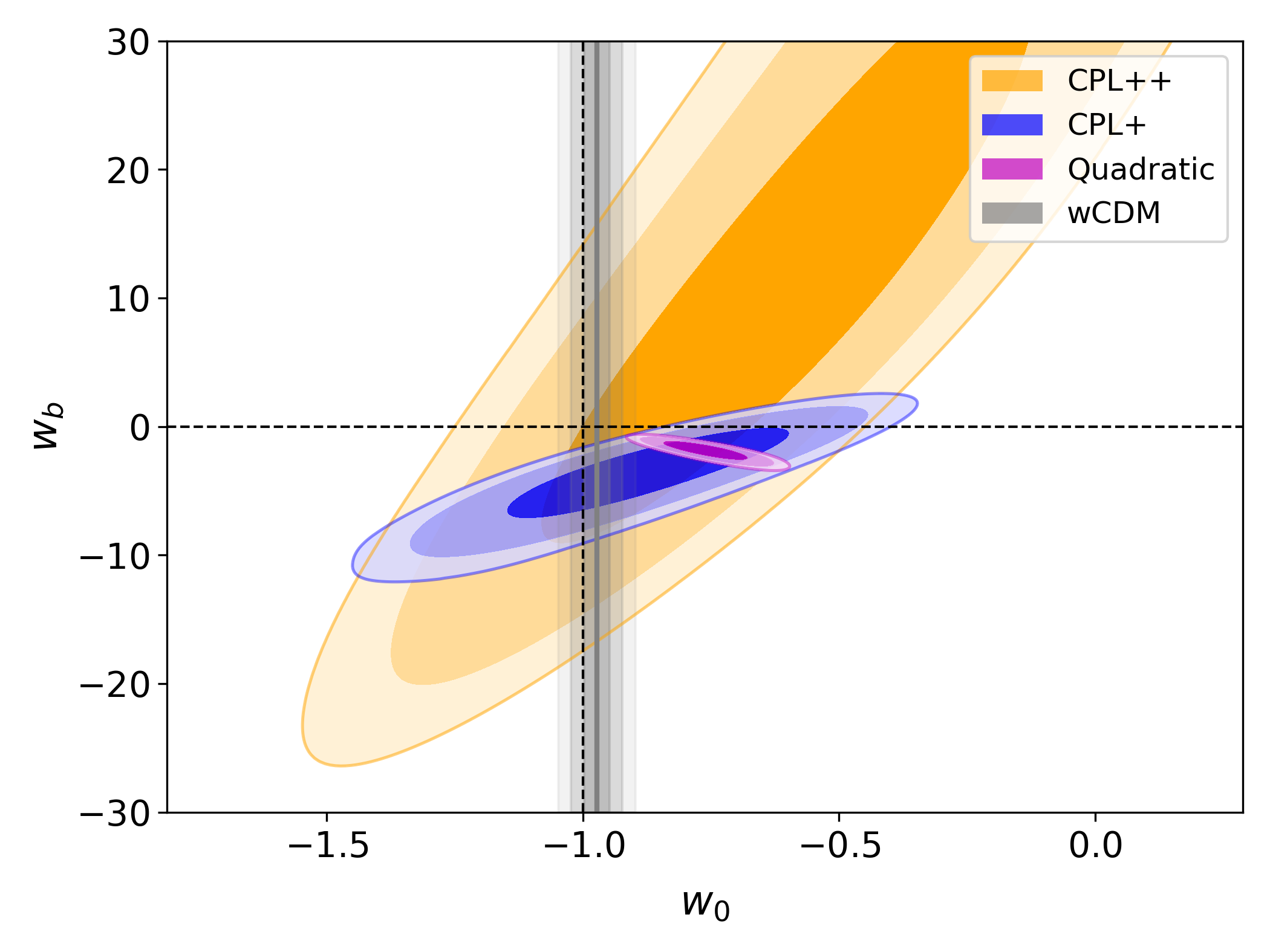}
    \includegraphics[width=0.5\textwidth]{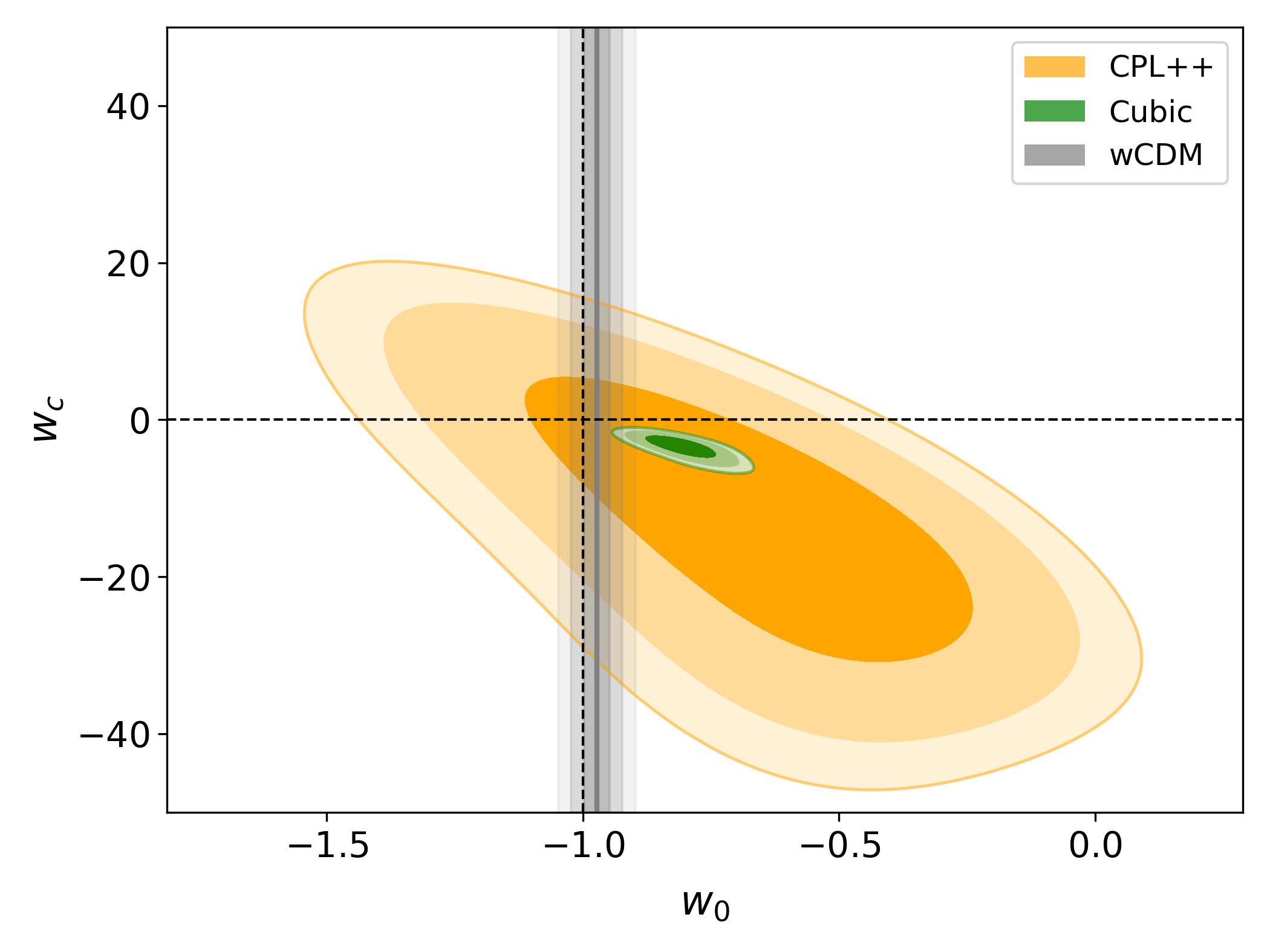}
    \caption{Same as Fig. \ref{fig:corner_pan}, But for CMB + DESI BAO (DR2) + Union3 combination.}
    \label{fig:corner_union}
\end{figure*}
In this appendix, we provide a visual representation of the observational constraints on the parameters of DE for the various parametrizations considered in our study. Here, we display the $1-3\sigma$ confidence contours in different two-dimensional parameter planes, complementing the numerical results presented in the main text.\\
In Fig. \ref{fig:corner_pan}, we plot the constraints obtained from the CMB + DESI BAO (DR2) + PantheonPlus combination. The contours visually represent the best-fit values and uncertainties reported in Table \ref{Tab:DE_parameters_pan}.\\
The top-left panel shows the constraints on the $w_0$-$w_a$ plane for the CPL, CPL$^+$, and CPL$^{++}$ parametrizations. Notice that $w_a$ is the linear parameter in the Taylor expansion of the equation of state and does not appear in the Quadratic and Cubic parameterizations. The top-right panel shows the constraints on the $w_0$-$w_b$ plane for the Quadratic, CPL$^+$, and CPL$^{++}$ parametrizations. Notice that standard CPL and Cubic parameterizations do not have the quadratic parameter $w_b$ in their definitions. Finally, the lower panel shows the constraints on the $w_0$-$w_c$ plane for CPL$^{++}$ and Cubic parametrizations. We note that other DE parameterizations under study do not have the cubic parameter $w_c$. In all panels, the vertical constraints on $w_0$ represent $1-3\sigma$ constraints on the wCDM cosmology with one constant parameter $w_0$. 
While the standard CPL model shows a clear deviation from $\Lambda$CDM at a high statistical significance level, the $w_0$-$w_a$ contours for CPL$^+$ and CPL$^{++}$ are much bigger and appear consistent with the $\Lambda$CDM point. Indeed, this artificial consistency is because of the projection effect, when we marginalize over the poorly constrained higher-order parameters ($w_b$ and $w_c$), their large uncertainties are propagated into the $w_0$ and $w_a$ parameters in two-dimensional contours. Furthermore, this misleading consistency is in  contrast to the full likelihood analysis reported in Table \ref{Tab:x2_pan}, where the MLE results indicate a moderately significant preference for evolving DE over the standard cosmology.\\
In Fig. \ref{fig:corner_union}, we present the two-dimensional contours using the CMB+DESI BAO (DR2)+Union3 combination. Notably, the preference for evolving DE is even more emphasized, as observed in the CPL contour being further apart from the $\Lambda$CDM point, indicating the stronger statistical deviation reported in Table \ref{Tab:x2_union}. In the case of CPL$^+$ and CPL$^{++}$, due to the projection effect described above, we have consistency with the $\Lambda$CDM cosmology in $w_0$-$w_a$ plane. While we observed these parameterizations have also significant deviations from the standard $\Lambda$CDM cosmology by the Likelihood statistical analysis in full space parameters.\newline
It is noted that since CPL$^+$ and CPL$^{++}$ include more free parameters, we expect the uncertainties on the DE parameters to be larger, given that the same dataset is used across all parameterizations under study. As a result, the errors associated with $w_b$ and $w_c$ propagate into the uncertainties of $w_0$ and $w_a$. Hence, the constraints on $w_0$ still cover the point $w_{\Lambda} = -1$, and also the constraints on $w_a$ cover the zero value in $w_0$-$w_a$ parameter space. However, the key point is that the best-fit value of $w_0$ in CPL$^+$ and CPL$^{++}$ parametrizations, compared to the $w_0w_a$ models, does not move closer to $-1$. Similarly, the best-fit value of $w_a$ does not shift closer to zero relative to the $w_0w_a$ parameterizations.

\end{document}